# Can artificial intelligence (AI) be used to accurately detect tuberculosis (TB) from chest X-rays? An evaluation of five AI products for TB triaging in a high TB burden setting

**Authors:**
Zhi Zhen Qin MSc[1], Shahriar Ahmed MHE[2], Mohammad Shahnewaz Sarker BSc[2], Kishor Paul MPH[2], Ahammad Shafiq Sikder Adel MPH[2], Tasneem Naheyan BEng[1], Rachael Barrett[1], Sayera Banu PhD[2(*)], Jacob Creswell PhD[1(*)].

**Affiliations:**
1. Stop TB Partnership, Chemin du Pommier 40, 1218 Le Grand-Saconnex, Geneva, Switzerland
2. International Centre for Diarrhoeal Disease Research, Bangladesh (icddr,b), Dhaka, Bangladesh
(*) Senior authors

**Correspondence to:**
Zhi Zhen Qin MSc
Stop TB Partnership, Chemin du Pommier 40, 1218 Le Grand-Saconnex
Geneva, Switzerland
Email zhizhenq@stoptb.org
Phone: + 41 79 540 75 22

# Summary

**Background**

Artificial intelligence (AI) products can be trained to recognize tuberculosis (TB)-related abnormalities on chest radiographs. Various AI products are available commercially, yet there is a lack of evidence on how their performance compares with each other and with radiologists.

**Objective**

We evaluated five AI software products for triaging TB using a large dataset that had not been used to train any commercial AI products.

**Methods**

Individuals (≥15 years old) presenting to three TB screening centers in Dhaka, Bangladesh, were recruited consecutively. Every participant was verbally screened for symptoms and received a chest x-ray (CXR) and an Xpert test. All CXR were read independently by a group of three Bangladeshi registered radiologists and five commercial AI products: CAD4TB (v7), InferRead®DR (v2), Lunit INSIGHT CXR (v4.9.0), JF CXR-1 (v2), and qXR (v3). We compared the performance of the AI products with each other, with the radiologists, and with the Target Product Profile (TPP) of triage tests. We used a new evaluation framework that simultaneously evaluates sensitivity, Xpert saving, and number needed to test to inform implementers' choice of vendor and threshold.

**Findings**

All five AI products significantly outperformed the radiologists. The areas under the receiver operating characteristic curve were qXR: 90·81% (95% CI:90·33-91·29%), CAD4TB: 90·34% (95% CI:89·81-90·87), Lunit INSIGHT CXR: 88·61% (95% CI:88·03%-89·20%), InferRead®DR: 84·90% (95% CI: 84·27-85·54%) and JF CXR-1: 84·89% (95% CI:84·26-85·53%). Only qXR and CAD4TB met the TPP with 74·3% and 72.9% specificity at 90% sensitivity respectively. Five AI algorithms reduced the number of Xpert tests required by 50% while maintaining a sensitivity above 90%. All AI algorithms performed worse among older age groups and people with prior TB history.

**Interpretation**

AI products can be highly accurate and useful triage tools for TB detection in high burden regions and outperform human readers.

**Funding**

Government of Canada

**Word count:** 298 words

## Introduction

The use of artificial intelligence (AI) technologies for medical diagnostics has accelerated rapidly in the past decade and AI-powered deep learning neural networks are increasingly being used to analyze medical images, such as chest radiographs or x-rays (CXR).[2,3]

CXR is recommended by the World Health Organization (WHO) as a screening and triage tool for tuberculosis (TB),[4] a disease which killed almost as many people worldwide in 2020 as COVID-19.[5] A triage test is used among people with TB symptoms and/or significant risk factors for TB[7]. The performance of CXR as a screening and triage tool has been limited by high inter-and intra-reader variability and moderate specificity,[4] as well as limited radiologist availability, especially in high TB-burden countries. Bangladesh is one such high burden country, with TB prevalence estimated at 260 per 100,000 population and greater prevalence in urban areas. [8]

AI technologies provide an opportunity to vastly increase image reading capacity in a variety of contexts. Such technology makes use of neural networks and deep learning to identify TB-related abnormalities from CXRs.[3] Inspired by the human nervous system, neural networks are interconnected functions, each comprised of a weight and a bias coefficient.[1] Through back-propagation, the networks "learn" by adjusting the weights and biases of the underlying functions based on the difference between predictions and ground truth in a training dataset.[1] Deep neural networks are structured in a number of layers, which increases the capacity of the machine to perform complex processes, such as parsing medical images.[6]

Several commercial AI products have emerged in recent years promising to identify TB-related abnormalities from digital CXR images.[9] AI algorithms produce a continuous abnormality score (from 0 to 100 or from 0 to 1) which represents the probability of the presence of the TB associated abnormalities .[10] Although some software comes with preset threshold scores, all products also allow users to customize the threshold score at any level to dichotomize the output into binary classification ("suggests confirmatory testing for TB" or not).[10]

In March 2021, the WHO updated the TB Screening Guidelines to recommend CAD software in place of human readers for analysis of digital CXR for TB screening and triage in individuals greater than 15 years old.[11] The WHO did not recommend specific products leaving many gaps for implementers to consider before a decision whether to implement, and which product, is made. Most available publications on AI feature earlier versions of one product and were conducted with the involvement of AI developers. Published evidence from impartial authors is therefore limited. [12-16] There is also a lack of sizeable external datasets to directly compare products.[15,17] Further, country programs and health professionals need performance measurements beyond accuracy, which is commonly reported as area under the Receiver Operating Characteristic (ROC)[18] curve (AUC)[15] (Annex 1), and guidance on operating point selection for different patient sources. To help implementers assess accuracy, we evaluated five AI software products for triaging TB using a large dataset that had not been used to train any commercial AI products. We also present a new analytical framework for selecting AI vendors and threshold scores in different settings.

## Methods

**Setting and test population**

This evaluation of AI software products to read CXR for TB followed the Standards for Reporting of Diagnostic Accuracy (STARD) Initiative on design and conduct of diagnostic accuracy evaluations .[19] In this retrospective study, we included all individuals (≥15 years old) who presented or were referred to three TB screening centers in Dhaka, Bangladesh (details, Annex 2) between May 15, 2014 and Oct 4, 2016.[20] Younger individuals were not included in the analysis as all evaluated AI products are only developed for those ≥15 years of age (Annex 3).

**Reading and testing Process**

After providing informed consent, each participant was verbally screened by healthcare workers for TB symptoms (procedure, Annex 2) using a standardized digital questionnaire and received a digital posterior-anterior (PA) CXR from a stationary Delft Easy DR X-ray System (machine specification and radiologist reading details in Annex 4). A few asymptomatic people suggested to have TB by the referring physicians also received a CXR. Three Bangladeshi radiologists, registered with Bangladesh Medical and Dental Council (BMDC) with 10, 6, and 1 year of experience (performing over 10,000 CXR reads a year minimum), worked part time for this project and alternated CXR reading. Blinded to any information except age and gender, the radiologists read 15-20 CXRs per day. They graded each CXR as normal or abnormal according to the TB prevalence survey handbook[21], and further classified abnormal CXRs into highly suggestive of TB, possibly TB, and abnormal but not TB, which could be analyzed separately.

The following AI companies agreed to participate in this independent study: CAD4TB (v7) by Delft Imaging Systems (Netherlands), InferRead®DR (v2) by Infervision (China), Lunit INSIGHT CXR for Chest Radiography (v4.9.0) by Lunit (South Korea), JF CXR-1 (v2) by JF Healthcare (China), and qXR (v3) by Qure.ai (India).[10] Detailed overviews of each AI product can be found in Annex 3. Each center was equipped with three 4-module GeneXpert systems and all individuals were asked to submit a fresh spot sputum sample for testing with the Xpert MTB/RIF (Xpert) assay. On average 12 Xpert tests were done per center daily. Xpert was repeated if the initial test failed (invalid, error, or no result). The final Xpert results were used as the bacteriological evidence and reference standard. All data collected was entered in a customized OpenMRS database and all CXR images were anonymized using a pydicom module[22] in python (script, Annex 5).

The five AI algorithms scored the anonymized images retrospectively, independently, and blinded to all information. A small sample of the anonymized CXRs were checked by the AI developers for image quality. No prior validation was done at the study site. We used CAD4TB cloud version to analyze the anonymized CXR files and installed the other four AI products on the Stop TB Partnership server to analyze the anonymized CXR files.

**Data analysis**

We first compared the performance of the group of three Bangladeshi radiologists with the five AI algorithms to detect bacteriologically positive (Bac+) TB-suggestive abnormalities. By dichotomizing the categories used by the Bangladeshi radiologists, we created three binary human reading classifications (A-C) varying the radiologist categories that were considered abnormal CXRs (Annex 6). To compare with the continuous output from AI, we calculated the sensitivity, specificity, positive predictive value (PPV), and negative predictive value (NPV) of the radiologists' three different binary classifications and the threshold score each AI product needed to match this sensitivity value for each one.[23] We then compared the difference in specificity, PPV, and NPV between human readings and those of the five AI algorithms using the McNemar test for paired proportions. We also compared each product against WHO's Target Product Profile (TPP) for a triage tool of ≥ 90% sensitivity and ≥70% specificity by altering threshold score to match each target value in turn and recording the performance.[6]

The AUCs were compared for each of the five algorithms using R programming (pROC package DeLong methods) for dependent AUCs.[24] Since ROC plots could mislead on the reliability of algorithm performance owing to an intuitive but wrong interpretation of the specificity in imbalanced datasets (low disease prevalence),[25] we also calculated the area under Precision-Recall (PRC) curve (PRAUC) for each (ROC and PRC methodology, Annex 7). Both ROC's and PRC's were generated for each product over a continuous range of threshold values.

Additionally, we assessed the distribution of abnormality scores disaggregated by Xpert results and prior history of TB. Finally, since the same threshold scores might provide different results in different populations, we evaluated the performance of the AI algorithms disaggregated by age, gender, prior TB history and patient sources using AUC and PRAUC.

**Evaluation framework**

We used an evaluation framework that analyzes performance beyond AUC, which is the standard approach of AI evaluations, to inform threshold selection by factoring in cost-effectiveness and the ability to triage. Products were evaluated in a hypothetical triage process whereby the AI score output would be used to triage all individuals in the study population for follow-on Xpert diagnosis based on a pre-defined threshold score. We calculated the proportion of subsequent Xpert assays saved (with 0% representing the Xpert testing-for-all scenario) as a proxy for a product's the cost-effectiveness. Likewise, the number of people needed to test (NNT) to find one Bac+ individual was used as a proxy for a product's ability to triage. We plotted the sensitivity against the proportion of Xpert saved to show the trade-off between finding as many Bac+ patients as possible and the cost savings of each AI algorithm. We produced visualizations of sensitivity, proportion of Xpert saved, and NNT over a continuous range of threshold scores in an evaluation framework to facilitate our understanding of threshold selection.

**Ethics**

All enrolled participants provided informed written consent (Annex 2). The study protocol was reviewed and approved by the Research Review Committee and the Ethical Review Committee at the International Centre for Diarrheal Disease Research, Bangladesh (icddr,b).

**Role of the AI developers**

The AI developers had no role in the study design, data collection, analysis plan, or writing of the study. The developers only had access to the CXR images and did not receive any information on the patients' demographic, symptom, medical, or testing data.

**Role of the funding source**

The funders of the study (Government of Canada) had no role in study design, data collection, data analysis, data interpretation, or writing of the report.

## Results

Between May 15, 2014 and Oct 4, 2016, a total of 24,009 individuals 15 years old and above visited the three TB centers and were enrolled for this study. Xpert tests needed to be repeated in 830 participants, of which, 15 remained invalid or showed an error after the second Xpert, and 24 did not have enough specimen for the second Xpert. Sixteen individuals did not have a valid or clear x-ray. After excluding these 55 individuals, a total of 23,954 (98·1%) individuals were included in the analysis (Figure 1). The median age was 42·0 [30·0, 57·0], 32·9% were female, and almost all (98·4%) reported at least one TB-related symptom. Reported symptoms included cough (89·9%), fever (79·6%), shortness of breath (54·4%), weight loss (62·8%), and hemoptysis (13·0%). The final sample included 3,586 (15·0%) participants who had a history of prior TB treatment. More than three quarters (n=17,541, 75·8%) of the participants were referred by public or private providers; 12.8% of participants were walk-ins; 10.7% referred by DOTS after a negative smear; while community screening and contact tracing made up just 142 (1.4%) of participants combined. The prevalence of Bac+ TB confirmed by Xpert was 15·3% overall (n=3,675), and 4·9% (n=181) of these cases were resistant to rifampicin, which is higher than the national average. About 14.3% of Xpert positive patient with prior history of TB had RIF resistance, and 3.1% TB patient without a history of TB had RIF resistance. The Bac+ rate and the RIF+ rate are both higher than the national average because of the high proportion of referrals and urban population in this study.[8] The radiologists graded 3,683 (15·4%) radiographs as Highly Suggestive of TB, 7,154 (29·9%) radiographs as Possibly TB, 3,625 (15.1%) radiographs as Abnormal-not TB, while 9,492 (39·6%) were read as normal (Table 1).

### Comparison between Radiologists' Reading and Prediction of the AI algorithms

The increase in the specificity of each AI algorithm when we selected threshold scores to match the sensitivity of the three radiological binary classifications are presented in Table 2. Classification A (highly suggestive of TB/normal) had a sensitivity of 38·9% (95%CI: 37·3% - 40·5%) and a specificity of 88·9% (95%CI: 88·5% - 89·4%). All AI algorithms had better specificity at the same sensitivity level and the improvement in specificity ranged from Lunit INSIGHT CXR's 9·02% (95%CI = 8·54-9·50%) to JF CXR-1's 4·55% (95%CI = 3·99-5·11%). The AI outperformed the human readers in classifications B and C as well. Additionally, all AI products significantly improved on human readers' PPV except InferRead DR, the increase in PPV was not statistically significant when compared with radiological Classification C with 0.8% (95% CI: -0·035% - 1·65%). However, for all three radiological binary classifications, the difference in NPVs between most AI products and human readers were not significant.

### Comparison with WHO TPP

Table 3 shows AI performance compared to the WHO TPP of ≥ 90% sensitivity and ≥70% specificity. At 90% sensitivity, algorithm specificities were 74·3% (qXR), 72·9% (CAD4TB), 67·2% (Lunit INSIGHT CXR), 62·1% (InferRead DR), 61·1% (JF CXR-1). At 70% specificity, algorithm sensitivities were 92·6% (qXR), 91·5% (CAD4TB), 88·8% (Lunit INSIGHT CXR), 85·0% (JF CXR-1), 84·0% (InferRead DR). Only qXR and CAD4TB met the TPP, with Lunit INSIGHT CXR coming close.

**Performance Comparison of the Five AI algorithms**

The trade-offs between sensitivity and specificity of the five AI algorithms can be visualized in the ROC (Figure 2-a) and precision-recall (Figure 2-b) curves. The AUCs of the ROC curve from high to low are qXR: 90·81% (95% CI:90·33-91·29%) and CAD4TB: 90·34% (95% CI:89·81-90·87), Lunit INSIGHT CXR: 88·61% (95% CI:88·03-89·20%), InferReadDR: 84·90% (95% CI:84·27-85·54%), and JF CXR-1: 84·89% (95% CI:84·26-85·53%). The AUCs of CAD4TB and qXR were significantly greater than the others. (p-value between $< 2.2e^{-16}$ and $1.0e^{-12}$, Annex 8). However, no significant difference in the AUCs was observed between JF CXR-1 and InferRead (p-value = 0.97, Annex 8). Above the 90% sensitivity mark, the ROC curves of all AI software were not significantly different. The PR curves and corresponding PRAUC scores (CAD4TB: 67·0%, qXR: 66·5%, Lunit INSIGHT CXR: 62·3%, JF CXR-1: 51·0% and InferReadDR: 49·6%) showed that some AI algorithms clearly had lower precision values for some given recall (sensitivity) values. Unlike ROC, it clearly shows a difference amongst the five AI algorithms. Example report outputs from AI products are provided in Annex 9.

**Evaluation Framework**

Figure 1-c shows that all five AI algorithms can reduce the number of Xpert tests required by 50%, while maintaining a sensitivity above 90%. However, as more diagnostic tests are saved (especially >60%), statistically significant differences in sensitivity become apparent between some of the algorithms. When Xpert testing is reduced by two thirds, the sensitivity across all algorithms ranged between 80% to 88%.

Although the performance of InferReadDR and JF CXR-1 could not be distinguished from the ROC and PR curves, the evaluation framework (Figure 2-d to Figure 2-f) showing the dynamics of sensitivity, proportion of Xpert saved and NNT with varying threshold scores, revealed significant differences in performance. For most of the decision thresholds (above approximately 0·15), JF CXR-1 had a higher sensitivity, but saved fewer Xpert tests and required higher NNT than InferReadDR (as shown in Figure 2-e/f). For instance, at a cut-off threshold at 0·8, JF CXR-1 is 93·0% (92·1-93·8%) sensitive, saves 48·7% of Xpert tests, and has an NNT of 3·6 (3·5-3·7). At the same threshold, InferReadDR is 35·4% (33·9-37·0%) sensitive, saves 90·5% of Xpert tests and has an NNT of 1·8 (1·7-1·8).

The performance of a single AI algorithm across the evaluation framework can be used to inform threshold selection. For most of threshold scores (0-0·9), the sensitivity JF CXR-1 remained above 90% (Figure 2-d), diagnostic Xpert tests saved remained between 30% and 60% (Figure 2-e) and the NNT was between 5 and 3 (Figure 2-f). The sensitivity of CAD4TB, Lunit INSIGHT CXR, and qXR remains above around 80% for most of the threshold scores (0-0·8), before quickly decreasing.

The results presented in Figure 2-d demonstrate that the threshold selection depends on the algorithm in question and the context in which it is being used. For example, the threshold required to achieve at least 90% sensitivity must be below 0·34 for InferReadDR, 0·50 for CAD4TB, below 0·60 for qXR and Lunit INSIGHT CXR, and below 0·93 for JF CXR-1.

**Density plot**

The stacked density plot in Annex 10 shows the distributions of the abnormality scores of the five AI algorithms disaggregated by Xpert outcomes and by prior TB history. The distributions of the five AI algorithms vary considerably, indicating different underlying neural networks and the effect that changing the threshold score can have for different products. Lunit INSIGHT CXR's, CAD4TB's, qXR's, and InferReadDR's density plot demonstrated good dichotomization pattern (between Bac+ and Bac-). Although almost all Bac+ participants received high abnormality scores (95-100) from JF CXR-1, so did many Bac- individuals. None of the distributions of the abnormality scores from the Bac- participants with prior TB history (the dark red bars) are left skewed.

**Subgroup Analysis**
All 5 AI algorithms showed significant variation in performance with age: performing worse in the older age group (>60 years old) than in both the younger (p values range $1.8e^{-16}$ [qXR] to $9.6e^{-8}$ [Lunit INSIGHT CXR]) and middle age group (p value range $6.7e^{-13}$ [qXR] to $6.6e^{-8}$ [Lunit INSIGHT CXR]). InferRead DR, CAD4TB, JF CXR-1, and qXR also performed significantly worse in the middle age group compared to the younger, although no significance was observed for the other products. All five AI algorithms performed significantly worse among people with prior TB history (p values from $1.6e^{-30}$ [InferRead DR] to $1.1e^{-12}$ [CAD4TB]). No significant differences were observed across genders, except JF-CXR-1 and Lunit INSIGHT CXR which performed better in males than females (p values= 0.0045 and 0.020 respectively) (AUC in Figure 3 and p-value in Annex 11).

AI performance also varied with patient source. All CAD performed significantly better among walk-ins than referrals from public and private facilities, and DOTS re-testing (p values from $7.5e^{-25}$ to 0.02), except Lunit INSIGHT CXR which showed no difference with public sector referral. qXR, JF CXR-1, and InferRead DR performed better among individuals from DOTS re-testing than private referrals (p = $2.1e^{-4}$, $3.0e^{-6}$, and $3.9e^{-3}$, respectively) and InferRead DR also performed significantly worse among individuals from community screening than walk-ins (p = 0.0065).

## Discussion

This is the largest independent study evaluating multiple AI algorithms as triage tests for TB with CXR, and the first published evaluation of JF CXR-1 and InferReadDR for detecting TB-suggested abnormalities, and of the latest version of CAD4TB (v7). Our study shows that the predictions made by the five algorithms significantly outperform experienced Bangladeshi human readers in detecting TB-related abnormalities.[14]

The AUCs indicate that all AI algorithms performed well, with qXR and CAD4TB being the two top performers. CAD4TB, Lunit INSIGHT CXR, and qXR's AUCs in this study are slightly lower than those in previous independent evaluations.[17,26] Our sub-analysis showed that CAD performance varied with demographic and clinical factors; as well as patient source; and therefore implied that variation exists in AI performance across different contexts and geographies. This cautions the generalization to other populations, particularly when selecting threshold scores for different populations. Training strategy and dataset (Annex 3) could explain these differences in CAD performance by affecting the ability of CAD to generalize learning to different populations. Further, the difference in performance of different CAD software is better visualized in PRC plot than in ROC curves, so should be utilized in future analyses with imbalanced datasets to inform vendor selection.

It is important that implementers can make informed decisions when selecting the threshold score specific to their settings. To do so, we used a new evaluation framework with important implementation-relevant indications, like confirmation tests saved and NNT, to measure the cost-effectiveness and the ability to triage. We observed that automated reading of CXR by all five AI products can keep sensitivity above 90% and at least halve the number follow-on diagnostic tests required. For large case finding programs that may have limited on-site Xpert testing capacity, there is a tradeoff between the number of cases identified and the proportion of Xpert that can be saved. A clear example of this is that choosing to save 70-80% of Xpert tests by allowing sensitivity to reduce to 70% misses 30% of TB cases. Additionally, the evaluation framework also indicates the difference in products that may have very similar performance using ROC and PRC alone.

The importance of using a more nuanced analytical framework for evaluation can be demonstrated by imagining different hypothetical case finding situations. For a program focused on capturing almost all people with TB and with access to many rapid diagnostic tests, to identify at least 95% of TB positive individuals: qXR would save the most confirmatory Xpert tests (54%), CAD4TB would save 51% Xpert tests, followed by JF CXR-1, Lunit INSIGHT CXR, and InferReadDR which would save 43%, 42%, and 41% of subsequent tests respectively. If we imagine another hypothetical case of a large active case finding program using CXR, but with a much more limited budget and the need to reduce the numbers of follow-on Xpert tests by 75% whilst accepting compromised sensitivity: qXR would have a sensitivity of 80·6% (79·2-81·8%) and CAD4TB with 79·7% (78·3-81·0%). This is followed by Lunit INSIGHT CXR with a sensitivity of 76·6% (75·1-77·9%), InferReadDR with 69·3% (67·7-70·8%) sensitivity, and JF CXR-1 with 68·5% (66·6%-69·7%) sensitivity. We recommend that the evaluation framework be included in future AI evaluations instead of only reporting on AUC.

The density plot shows that underlying neural networks of the five AI algorithms were constructed very differently and that no universal threshold score can be applicable to all AI algorithms. Moreover, the density plots of the Bac- individuals with prior TB history indicates the algorithms' poor ability to differentiate between old scarring and active lesions could lead to excessive recall in this group. We hypothesize that the abnormalities on the chest due to age and prior TB history influenced the classification of active TB. The overall performance of the five AI algorithms differed between age groups, patient sources, and prior TB history, although performed similarly across genders (Figure 3, Annex 11). Interestingly, accuracy (i.e. AUC) was lowest among those who presented themselves and were recruited through community-based case finding. This implies that the threshold scores likely need to be different depending on the population tested and with sub populations. We further recommended the manufacturers to include basic demographic and clinical data to further improve their AI products in future software iterations.

Our results document the performance of five products at one point in time. New products are set to emerge in the near future and updated software versions are launched almost annually.[10] Two products in this evaluation had not been previously evaluated in peer-reviewed journals. Unlike traditional diagnostic tests which take years to produce and update, the performance of AI improves incredibly fast. Future guidance from bodies like WHO must prepare for this speed of change and independent evaluation libraries are required to help implementers understand the latest performance.

**Limitations**

Due to logistic and budgetary constraints, we did not use culture as the reference standard, meaning that some people with Xpert-negative, culture-positive TB have incorrectly been labelled as not having TB. We also did not have access to Xpert Ultra, which is more sensitive than Xpert, in Bangladesh during the study. Due to limited number of asymptomatic individuals, we did not stratify by symptoms and perform subgroup analysis. We did not conduct HIV testing because Bangladesh has a low HIV prevalence.[5] However, the performance among different sub-populations, especially those living with HIV who often present with atypical radiological images, needs to be better documented.[27] Similarly, we excluded children from our study population, even though some (but not all) of the products included are licensed for use in younger age groups. [27,27] Such decisions, not to evaluate performance in children and HIV+ individuals, limits the generalizability of our findings. Further evaluation of CAD in children is necessary. Each CXR was read by one Bangladeshi radiologist, not by a panel of radiologists. However, the intended use of these AI algorithms is in resource-constrained settings with few or no radiologists and neither resources nor time permitted multiple readings of high numbers of images. Further, human readers were blinded to clinical and demographic information except age and gender, though in the field reading of CXR could be informed by this data. Additionally, the analysis includes only three human reads as a comparison point. We caution against extrapolating the study findings to rural areas as our study was done in metropolitan Dhaka where experienced human readers are often more available. CAD software may perform even better compared with the radiologists in rural and low resourced areas. Additionally, only one brand of x-ray machine was used in this study due to procurement constraints. Lastly, we did not conduct this study prospectively and did not collect implementation data such as programmatic costs, setup, services, user experience, etc.

**Conclusions**

Our results demonstrate that all five AI algorithms outperformed experienced certified radiologists and could save follow-on Xpert testing and reduce NNT whilst maintaining high sensitivity. ROC and precision recall curves are powerful tools for evaluation, however, additional metrics and analysis, including our new evaluation framework of sensitivity, confirmation tests saved, and NNT with varying threshold scores, will help implementers with threshold and software selection.

**Author Contributions**

The study was conceived by ZZQ, JC and SB. Data collection was led by SB, KP, SA, MS. Data cleaning and verification was done by ZZQ, SB, KP, SA, data analysis and interpretation conducted by ZZQ, TN and JC; ZZQ wrote the first draft of the manuscript. ZZQ, JC and RB revised the manuscript. All authors contributed to and approved the final manuscript.

**Declaration of interests**

None declared.

**Data sharing**

The datasets used in this study can be available upon reasonable request to the corresponding author. CXR images will not be provided as these are withheld by the corresponding author's organization to reserve their use for product evaluations.

**Acknowledgments**
Funding support of this project came from Global Affairs Canada through the Stop TB Partnership's TB REACH initiative (Grant no: STBP/TBREACH/GSA/W5-24). ZZQ, JC, TN and RB work for the Stop TB Partnership. We were allowed to use the AI algorithm free of charge by all 5 AI companies for research purposes, but the companies had no influence over the research question, nor any other aspect of the work carried out, nor any impact on the transparency of the Article.

# Tables

*Table 1. Characteristics of the 23,954 individuals included in this study.*

| | Overall N = 23,954 | Xpert Results | | | TB History | | | Radiologist grading | | | | |
|---|---|---|---|---|---|---|---|---|---|---|---|---|
| | | **Xpert Positive N = 3,675** | **Xpert Negative N = 20,279** | **p test** | **Previous TB Cases N=3,586** | **New Case N = 20,341** | **p test** | **Abnormal-highly suggestive of TB N= 3,683** | **Abnormal -Possibly TB N= 7,154** | **Abnormal-not TB N=3,625** | **Normal N= 9,492** | **p test** |
| **Age (median [IQR])** | 42.0 [30.0, 57.0] | 37.0 [27.0, 53.0] | 43.0 [31.0, 58.0] | <0.001 | 44.0 [31.0, 58.0] | 42.0 [30.0, 57.0] | <0.001 | 45.0 [31.0, 60.0] | 47.0 [32.0, 62.0] | 54.0 [39.0, 65.0] | 35.0 [28.0, 48.0] | <0.001 |
| **Age Group (%)** | | | | <0.001 | | | <0.001 | | | | | <0.001 |
| Young age (15-25 years) | 2666 (11.1) | 664 (18.1) | 2002 (9.9) | | 309 (8.6) | 2355 (11.6) | | 377 (10.2) | 708 (9.9) | 182 (5.0) | 1399 (14.7) | |
| Middle age (25-60 years) | 16056 (67.0) | 2378 (64.7) | 13678 (67.4) | | 2437 (68.0) | 13606 (66.9) | | 2363 (64.2) | 4488 (62.7) | 2048 (56.5) | 7157 (75.4) | |
| Old age (⩾ 60 years) | 5232 (21.8) | 633 (17.2) | 4599 (22.7) | | 840 (23.4) | 4380 (21.5) | | 943 (25.6) | 1958 (27.4) | 1395 (38.5) | 936 (9.9) | |
| **Gender = Female (%)** | 7876/16078 (32.9/67.1) | 1061/2614 (28.9/71.1) | 6815/13464 (33.6/66.4) | <0.001 | 1163/2423 (32.4/67.6) | 6706/13635 (33.0/67.0) | 0.541 | 1057/2626 (28.7/71.3) | 2196/4958 (30.7/69.3) | 1337/2288 (36.9/63.1) | 3286/6206 (34.6/65.4) | <0.001 |
| **TB Medication History = Yes (%)** | 3586 (15.0) | 608 (16.6) | 2978 (14.7) | 0.004 | 3586 (100.0) | 0 (0.0) | <0.001 | 969 (26.3) | 1467 (20.5) | 396 (10.9) | 754 (8.0) | <0.001 |
| **Symptoms** | | | | | | | | | | | | |
| Cough = Yes (%) | 21494 (89.9) | 3416 (93.1) | 18078 (89.3) | <0.001 | 3176 (88.6) | 18318 (90.1) | 0.009 | 3376 (91.7) | 6443 (90.2) | 3331 (92.0) | 8344 (88.0) | <0.001 |
| Fever = Yes (%) | 19041 (79.6) | 3200 (87.2) | 15841 (78.2) | <0.001 | 2931 (81.7) | 16110 (79.2) | 0.001 | 3198 (86.9) | 5951 (83.3) | 2825 (78.0) | 7067 (74.5) | <0.001 |
| Short of breath = Yes (%) | 13011 (54.4) | 2062 (56.2) | 10949 (54.1) | 0.022 | 1888 (52.7) | 11123 (54.7) | 0.024 | 2287 (62.2) | 4064 (56.9) | 1953 (54.0) | 4707 (49.7) | <0.001 |
| Weight Loss = Yes (%) | 15035 (62.8) | 2774 (75.6) | 12261 (60.5) | <0.001 | 2475 (69.0) | 12560 (61.7) | <0.001 | 2871 (78.0) | 4964 (69.5) | 2201 (60.8) | 4999 (52.7) | <0.001 |
| Hemoptysis = Yes (%) | 3103 (13.0) | 471 (12.8) | 2632 (13.0) | 0.807 | 511 (14.2) | 2592 (12.7) | 0.014 | 597 (16.2) | 892 (12.5) | 452 (12.5) | 1162 (12.3) | <0.001 |
| Any symptom(s) = Yes (%) | 23582 (98.4) | 3644 (99.2) | 19938 (98.3) | <0.001 | 3499 (97.6) | 20056 (98.6) | <0.001 | 3652 (99.2) | 7039 (98.4) | 3581 (98.8) | 9310 (98.1) | <0.001 |
| **Patient Source (%)** | | | | <0.001 | | | <0.001 | | | | | <0.001 |
| Community Screening | 170 (0.7) | 16 (0.4) | 154 (0.8) | | 20 (0.6) | 150 (0.8) | | 28 (0.8) | 30 (0.4) | 15 (0.4) | 97 (1.0) | |
| Contacts | 172 (0.7) | 9 (0.3) | 163 (0.8) | | 11 (0.3) | 161 (0.8) | | 12 (0.3) | 31 (0.4) | 20 (0.6) | 109 (1.2) | |
| Private Referral | 17056 (73.0) | 2822 (78.7) | 14234 (71.9) | | 2515 (72.2) | 14521 (73.1) | | 2796 (77.8) | 5565 (79.3) | 2647 (76.3) | 6048 (65.1) | |
| Public DOTS Retesting | 2496 (10.7) | 436 (12.2) | 2060 (10.4) | | 597 (17.1) | 1897 (9.6) | | 510 (14.2) | 685 (9.8) | 282 (8.1) | 1019 (11.0) | |
| Public Referral | 485 (2.1) | 100 (2.8) | 385 (1.9) | | 47 (1.3) | 438 (2.2) | | 39 (1.1) | 212 (3.0) | 94 (2.7) | 140 (1.5) | |
| Walk-in | 2992 (12.8) | 204 (5.7) | 2788 (14.1) | | 294 (8.4) | 2694 (13.6) | | 207 (5.8) | 492 (7.0) | 409 (11.8) | 1884 (20.3) | |
| **Bac positive (Xpert) (%)** | 3675 (15.3) | 3675 (100.0) | 0 (0.0) | <0.001 | 608 (17.0) | 3063 (15.1) | 0.004 | 1441 (39.1) | 1813 (25.3) | 240 (6.6) | 181 (1.9) | <0.001 |
| **MTB Burden (%)** | | | | <0.001 | | | <0.001 | | | | | <0.001 |
| Very Low | 634 (17.3) | 634 (17.3) | | | 127 (20.9) | 507 (16.5) | | 192 (13.3) | 304 (16.7) | 60 (25.1) | 79 (43.6) | |
| Low | 1093 (29.7) | 1093 (29.7) | | | 192 (31.5) | 903 (29.4) | | 391 (27.1) | 570 (31.4) | 82 (34.3) | 52 (28.7) | |
| Medium | 1299 (35.4) | 1299 (35.4) | | | 187 (30.7) | 1111 (36.2) | | 547 (37.9) | 643 (35.4) | 69 (28.9) | 41 (22.7) | |
| High | 648 (17.6) | 648 (17.6) | | | 103 (16.9) | 546 (17.8) | | 312 (21.6) | 301 (16.6) | 28 (11.7) | 9 (5) | |
| **RIF Result (%)** | | | | <0.001 | | | <0.001 | | | | | <0.001 |
| Detected | 181 (4.9) | 181 (4.9) | | | 87 (14.3) | 94 (3.1) | | 91 (6.3) | 79 (4.4) | 8 (3.3) | 3 (1.7) | |
| Not Detected | 3475 (94.7) | 3475 (94.7) | | | 520 (85.5) | 2951 (96.5) | | 1347 (93.5) | 1724 (95.2) | 230 (95.8) | 174 (96.7) | |
| Indeterminate | 14 (0.4) | 14 (0.4) | | | 1 (0.2) | 13 (0.4) | | 1 (0.1) | 8 (0.4) | 2 (0.8) | 3 (1.7) | |
| **Radiologist grading (%)** | | | | <0.001 | | | <0.001 | | | | | <0.001 |
| Abnormal-Highly suggestive | 3683 (15.4) | 1441 (39.2) | 2242 (11.1) | | 969 (27.0) | 2712 (13.3) | | | | | | |
| Abnormal-Possibly TB | 7154 (29.9) | 1813 (49.3) | 5341 (26.3) | | 1467 (40.9) | 5679 (27.9) | | | | | | |
| Abnormal-not TB | 3625 (15.1) | 240 (6.5) | 3385 (16.7) | | 396 (11.0) | 3224 (15.8) | | | | | | |
| Normal | 9492 (39.6) | 181 (4.9) | 9311 (45.9) | | 754 (21.0) | 8726 (42.9) | | | | | | |
| **AI abnormality scores (median [IQR])** | | | | | | | | | | | | |
| CAD4TB | 17.4 [2.4, 80.1] | 97.1 [88.0, 99.2] | 9.0 [1.8, 54.5] | <0.001 | 61.2 [16.9, 91.1] | 11.6 [2.0, 74.3] | <0.001 | 90.5 [64.5, 98.1] | 71.5 [33.8, 94.6] | 10.8 [2.6, 44.8] | 2.2 [0.8, 6.4] | <0.001 |
| qXR | 24.0 [3.0, 78.0] | 89.0 [82.0, 93.0] | 11.0 [2.0, 61.0] | <0.001 | 68.0 [25.0, 85.0] | 15.0 [3.0, 75.0] | <0.001 | 85.0 [74.0, 91.0] | 72.0 [41.0, 86.0] | 16.0 [4.0, 51.0] | 3.0 [2.0, 7.0] | <0.001 |
| Lunit INSIGHT CXR | 29.0 [2.0, 86.0] | 95.0 [88.0, 97.0] | 10.0 [2.0, 76.0] | <0.001 | 81.0 [31.0, 91.0] | 15.0 [2.0, 84.0] | <0.001 | 91.0 [83.0, 96.0] | 82.0 [51.0, 93.0] | 19.0 [3.0, 64.0] | 2.0 [1.0, 5.0] | <0.001 |
| JF CXR-1 | 85.0 [8.1, 99.8] | 100.0 [99.6, 100.0] | 58.4 [5.3, 99.3] | <0.001 | 99.6 [88.1, 100.0] | 69.5 [5.9, 99.7] | <0.001 | 100.0 [99.7, 100.0] | 99.6 [95.6, 100.0] | 71.6 [21.3, 97.5] | 6.0 [1.6, 31.8] | <0.001 |
| InferReadDR | 28.2 [13.6, 64.9] | 74.9 [59.4, 83.1] | 22.1 [12.5, 54.0] | <0.001 | 59.7 [31.2, 74.6] | 23.8 [12.9, 60.6] | <0.001 | 72.1 [59.3, 81.0] | 58.9 [35.1, 74.8] | 23.4 [15.1, 42.8] | 13.3 [10.0, 19.7] | <0.001 |

*Table 2 Comparison of sensitivity and specificity between radiologists' reading and the predictions of the AI algorithms*

| | **Bangladeshi Radiologists** | | | | **Commercial AI Products** | | | | | **Differenceł between AI and radiologists reading** | | |
|---|---|---|---|---|---|---|---|---|---|---|---|---|
| **Human Binary classification** | **Sensitivity** | **Specificity** | **PPV** | **NPV** | **Product** | **Threshold Score** | **Specificity (95%CI)** | **PPV** | **NPV** | **Specificity (95%CI)** | **PPV (95%CI)** | **NPV (95%CI)** |
| Binary classification A | 38.9% (37.3-40.5%) | 88.9% (88.5-89.4%) | 39.1%, (37.5-40.7%) | 89.0%, (88.5-89.4%) | CAD4TB | 98 | 97.8% (97.6-98.0%) | 76.2% (74.2-78.1%) | 89.8% (89.4-90.2%) | 8.9% (8.40-9.37%) | 37.1% (36.2-38.0%) | 0.79% (0.189-1.40%) |
| | | | | | InferReadDR | 0.79 | 94.2% (93.8-94.5%) | 54.9% (52.9-56.8%) | 89.4% (89.0-89.8%) | 5.23% (4.68-5.78%) | 15.7% (14.8-16.7%) | 0.44% (-0.169-1.05%) |
| | | | | | JF CXR-1 | 1.00 | 93.5% (93.1-93.8%) | 54.2% (52.4-56.0%) | 89.9% (89.5-90.3%) | 4.55% (3.99-5.11%) | 15.1% (14.1-16.0%) | 0.96% (0.356-1.56%) |
| | | | | | Lunit INSIGHT CXR | 0.96 | 98.0% (97.8-98.1%) | 75.5% (73.3-77.5%) | 89.1% (88.7-89.5%) | 9.02% (8.54-9.50%) | 36.3% (35.4-37.2%) | 0.15% (-0.465-0.76%) |
| | | | | | qXR | 0.91 | 97.9% (97.7-98.1%) | 75.9% (73.8-77.8%) | 89.5% (89.1-89.9%) | 8.9% (8.45-9.42%) | 36.8% (35.9-37.7%) | 0.53% (-0.0833-1.13%) |
| Binary classification B | 88.5% (87.4-89.5%) | 62.5% (61.8-63.1%) | 30.0%, (29.2-30.9%) | 96.8%, (96.5-97.1%) | CAD4TB | 57 | 75.8% (75.2-76.4%) | 40.0% (39.0-41.1%) | 97.3% (97.0-97.5%) | 13.4% (12.5-14.3%) | 10% (9.07-10.9%) | 0.51% (0.175-0.844%) |
| | | | | | InferReadDR | 0.37 | 64.5% (63.8-65.1%) | 31.2% (30.3-32.1%) | 96.8% (96.5-97.1%) | 2.01% (1.06-2.96%) | 1.2% (0.286-2.09%) | 0.05% (-0.301-0.393%) |
| | | | | | JF CXR-1 | 0.95 | 64.1% (63.4-64.7%) | 31.0% (30.1-31.9%) | 96.8% (96.5-97.1%) | 1.62% (0.66-2.57%) | 1% (0.0709-1.87%) | 0.03% (-0.318-0.376%) |
| | | | | | Lunit INSIGHT CXR | 0.66 | 70.3% (69.7-71.0%) | 35.3% (34.3-36.3%) | 97.1% (96.8-97.4%) | 7.87% (6.95-8.80%) | 5.3% (4.36-6.19%) | 0.31% (-0.0319-0.648%) |
| | | | | | qXR | 0.64 | 76.7% (76.1-77.2%) | 40.9% (39.8-41.9%) | 97.4% (97.1-97.6%) | 14.2% (13.3-15.1%) | 10.8% (9.89-11.8%) | 0.56% (0.228-0.894%) |
| Binary classification C | 95.0% (94.3-95.7%) | 45.7% (45.0-46.4%) | 24.2%, (23.5-24.9%) | 98.1%, (97.8-98.4%) | CAD4TB | 18 | 58.5% (57.8-59.1%) | 29.5% (28.7-30.3%) | 98.5% (98.2-98.7%) | 12.8% (11.9-13.8%) | 5.3% (4.46-6.20%) | 0.39% (0.131-0.646%) |
| | | | | | InferReadDR | 0.20 | 47.5% (46.8-48.2%) | 25.0% (24.2-25.7%) | 98.1% (97.9-98.4%) | 1.80% (0.82-2.78%) | 0.8% (-0.035-1.65%) | 0.06% (-0.213-0.325%) |
| | | | | | JF CXR-1 | 0.53 | 49.0% (48.3-49.7%) | 25.5% (24.8-26.2%) | 98.2% (97.9-98.5%) | 3.31% (2.32-4.29%) | 1.3% (0.481-2.17%) | 0.11% (-0.159-0.376%) |
| | | | | | Lunit INSIGHT CXR | 0.07 | 47.8% (47.1-48.5%) | 25.6% (24.8-26.3%) | 98.2% (97.9-98.5%) | 2.16% (1.17-3.14%) | 1.4% (0.546-2.24%) | 0.12% (-0.152-0.382%) |
| | | | | | qXR | 0.35 | 63.5% (62.9-64.2%) | 32.2% (31.3-33.1%) | 98.6% (98.4-98.8%) | 17.9% (16.9-18.8%) | 8% (7.16-8.92%) | 0.48% (0.230-0.738%) |

ɫ a positive difference in specificity, PPV, or NPV means that the specificity, PPV, or NPV of AI algorithm is higher than the that of the Bangladeshi radiologists, when matching sensitivity. A negative difference in specificity, PPV, or NPV means that the specificity, PPV, or NPV of AI algorithm is lower than the that of the Bangladeshi radiologists, when matching sensitivity. For example, the difference in specificity = the specificity of an AI product – the specificity of the corresponding radiological binary classification.

Table 3 Comparison of CAD products against FIND's TPP when matching either sensitivity or specificity

| AI | Score | Sensitivity | Specificity |
|---|---|---|---|
| **Fixing Sensitivity to 90%** | | | |
| **CAD4TB** | 0.50 | 90.0% (89.0-91.0%) | 72.9% (72.3-73.5%) |
| **InferRead DR** | 0.34 | 90.3% (89.3-91.3%) | 62.1% (61.4-62.7%) |
| **JF-CXR-1** | 0.92 | 90.4% (89.4-91.3%) | 61.1% (60.4-61.8%) |
| **Lunit INSIGHT CXR** | 0.6 | 90.1% (89.0-91.0%) | 67.2% (66.6-67.9%) |
| **qXR** | 0.6 | 90.2% (89.2-91.1%) | 74.3% (73.3-74.9%) |
| **Fixing Specificity to 70%** | | | |
| **CAD4TB** | 0.44 | 91.5% (90.5-92.4%) | 70.0% (69.4-70.6%) |
| **InferRead D** | 0.47 | 84.0% (82.8-85.2%) | 70.6% (69.9-71.2%) |
| **JF-CXR-1** | 0.98 | 85.0% (83.8-86.2%) | 68.8% (68.2-69.5%) |
| **Lunit INSIGHT CXR** | 0.67 | 88.8% (87.7-89.8%) | 70.1% (69.4-70.7%) |
| **qXR** | 0.51 | 92.6% (91.7-93.4%) | 70.3% (69.6-70.9%) |

Marks the closest available match to a specificity value of 70%.

## Figures

*Figure 1. The diagnostic process used by the 3 TB screening sites in the study*

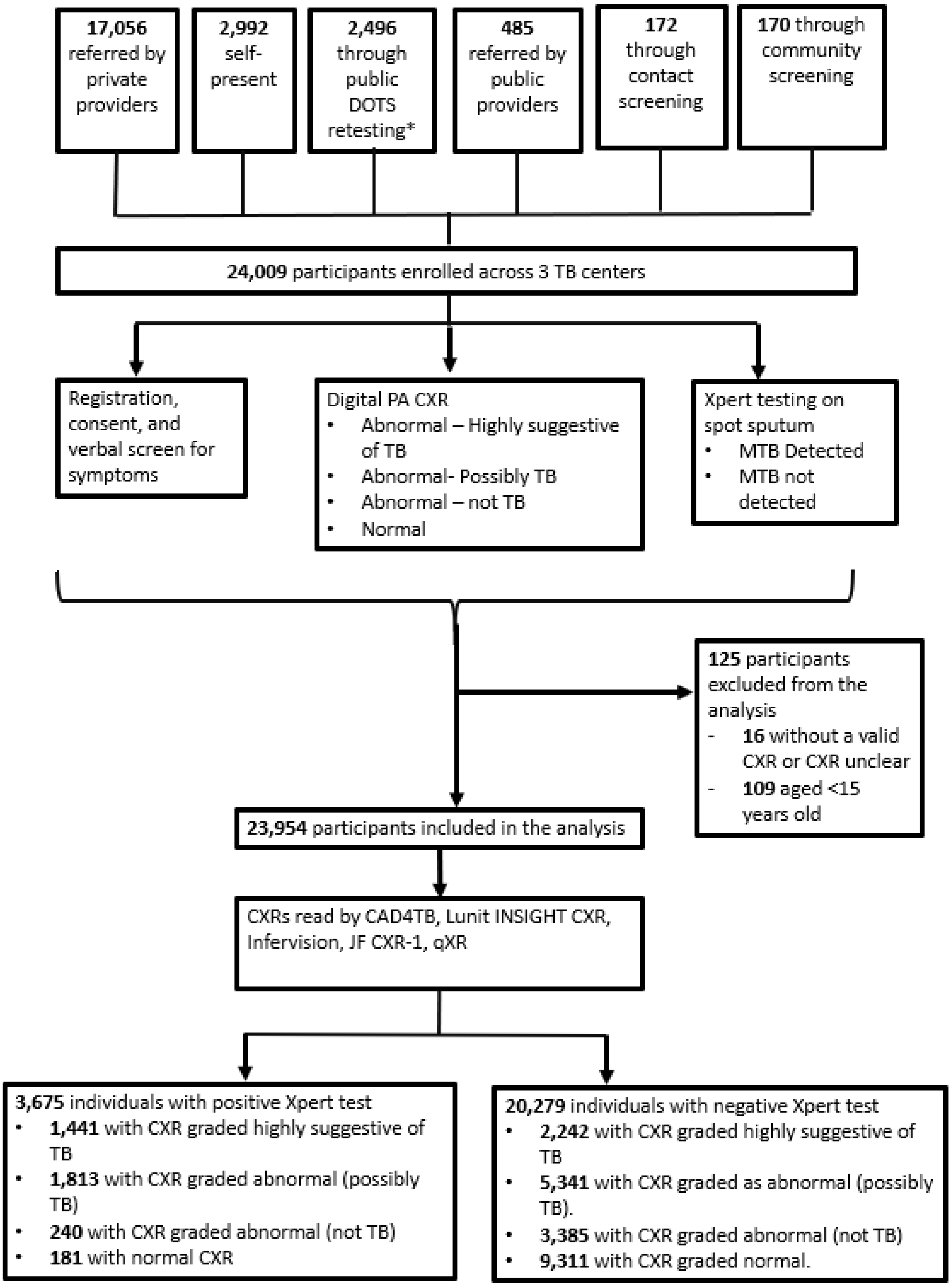

* DOTS (Directly Observed Treatment Short course) centers are specialized facilities for the diagnosis and treatment of TB patients. DOTS re-testing refers to individuals referred from DOTS center after a negative smear.

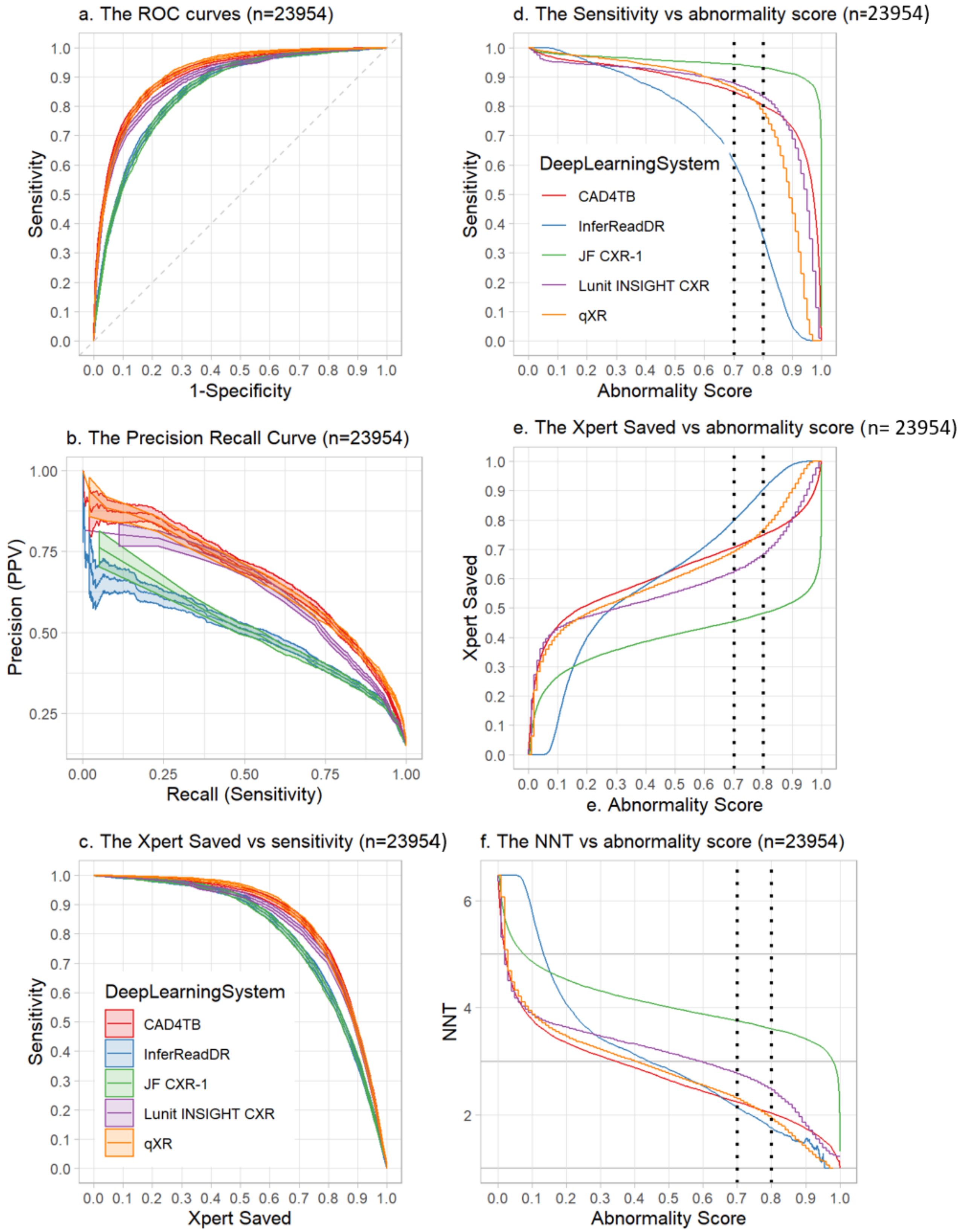

Figure 2. A) ROC curve for 5 AI products over continuous range of threshold scores (n=23,954). B) PRC curve for 5 AI products over continuous range of threshold scores (n=23,954). C) Tradeoff between sensitivity and proportion of subsequent Xpert tests saved (n=23,954). D) Sensitivity of 5 AI products over a continuous range of threshold scores (n=23,954). E) Proportion of subsequent Xpert tests saved over a continuous range of threshold scores (n=23,954). F) Number needed to test (NNT) over a continuous range of threshold scores (n=23,954)

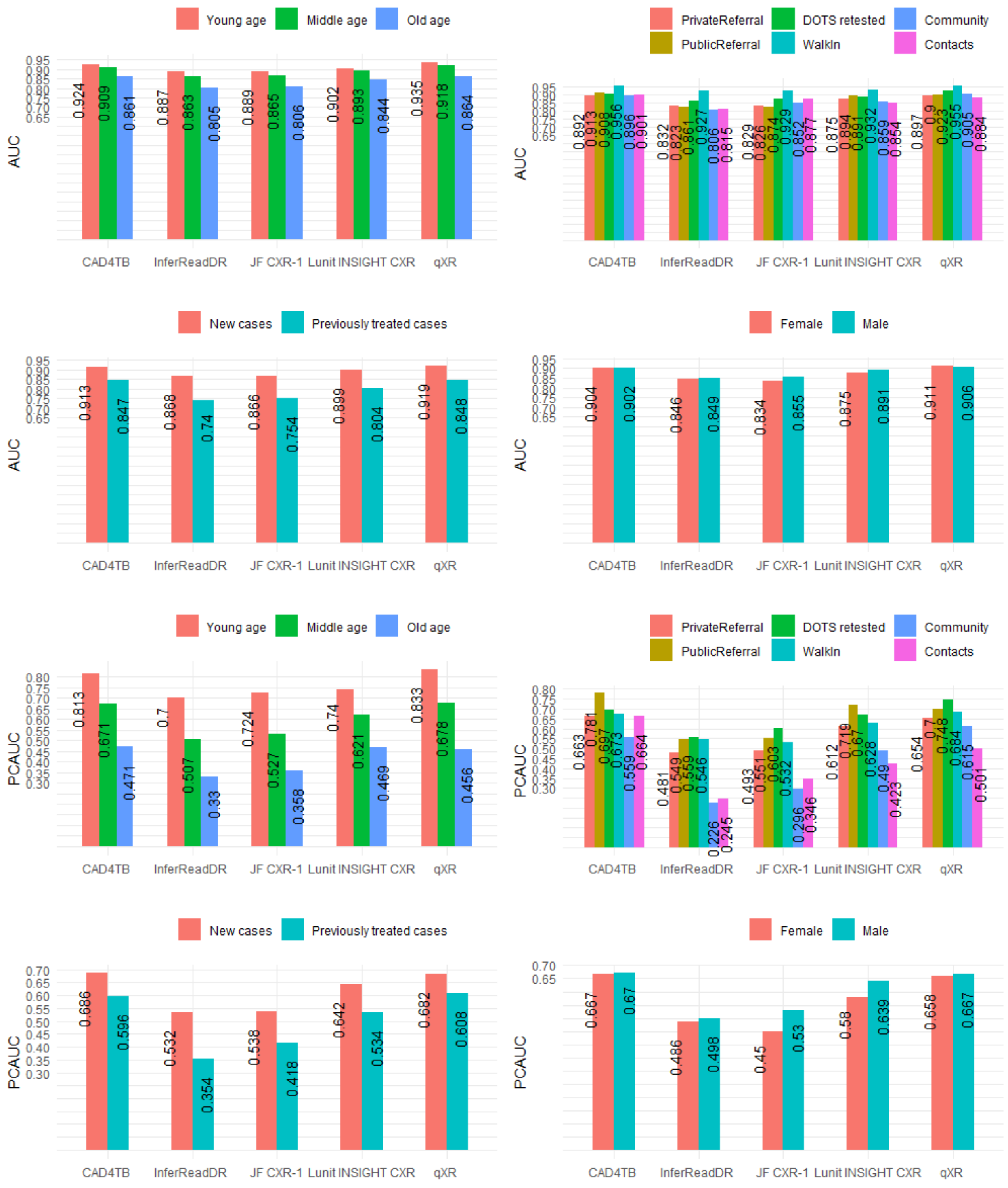

# Can artificial intelligence (AI) be used to accurately detect tuberculosis (TB) from chest X-rays? An evaluation of five AI products for TB triaging in a high TB burden setting

Figure 3. The AUC and PRAUC stratified into subgroups.
The first two rows are AUC of ROC curves and the third and fourth row is the AUC of PR curves. For each AUC/PRAUC is disaggregated by age (young age (15-25 years old), middle age (25-60 years old), older age (above 60 years old)); patient source; TB history and gender, respectively,

Annexes

**Authors:**
Zhi Zhen Qin MSc[1], Shahriar Ahmed MHE[2], Mohammad Shahnewaz Sarker BSc[2], Kishor Paul MPH[2], Ahammad Shafiq Sikder Adel MPH[2], Tsasneem Naheyan BEng[1], Rachael Barrett[1], Sayera Banu PhD[2(*)], Jacob Creswell PhD[1(*)].

**Affiliations:**
1. Stop TB Partnership, Chemin du Pommier 40, 1218 Le Grand-Saconnex, Geneva, Switzerland
2. International Centre for Diarrhoeal Disease Research, Bangladesh (icddr,b), Dhaka, Bangladesh
(*) Senior authors

## Annex 1: AI Software prior product performance against Xpert reference standard

| AI Product | Version evaluated | Reported AUC of ROC curve* (95% CI) | Reference |
|---|---|---|---|
| CAD4TB | 6 | 0.92 (0.9-0.95) | Qin 2019 |
| | 6 | 0.885 (0.874-0.894) | Murphy 2020 |
| | 5 | 0.865 (0.853-0.876) | |
| | 4 | 0.878 (0.865-0.889) | |
| | 3 | 0.794 (0.781-0.807) | |
| | 3.07 | 0.79 (0.78-0.81) | Zaidi 2018 |
| | 3.07 | 0.74 (0.73-0.75) | Rahman 2017 |
| | 1.08 | 0.73 (0.64-0.80) | Maduskar 2013 |
| Lunit INSIGHT | 4.7.2 | 0.94 (0.93-0.96) | Qin 2019 |
| qXR | 2 | 0.94 (0.92-0.97) | Qin 2019 |

* AUC = area under the curve; ROC curve = Receiver operating characteristic curve
There is no reported data for InferRead DR Chest or JF CXR-1, or version 7 of CAD4TB.

## Annex 2 Description of TB Screening Centers and consent procedure

Participants were recruited from individuals presenting or referred to three centers in metropolitan Dhaka: icddr,b Mohakhali Tuberculosis Screening & Treatment Center (TBSTC), icddr,b Golapbagh TBSTC, and icddr,b Dhanmondi TBSTC. Everyone was screened TB symptoms (any cough, fever, shortness of breath, weight loss, hemoptysis), and then received chest x-ray and GeneXpert. How patients arrived at the TB screening center was documented (whether this was referral by public or private providers, walk-in, referral by DOTS center, from community screening or contact tracing) and individuals were asked to self-report history of TB.

For literate adults (>17 years) consent was taken in the form of a signature. For children and adolescents (15-17 years) we obtained consent from both the parent/guardian and child (assent). For illiterate person the consent was read in front of the person to confirm that he/she understands his/ her engagements with the study in the presence of a witness. They consent by their thumb print in front of a witness.

## Annex 3: AI software specification

| Country | Netherlands | China | China | South Korea | India |
|---|---|---|---|---|---|
| **Product** | **CAD4TB** | **InferRead DR Chest** | **JF CXR-1** | **Lunit Insight CXR** | **qXR** |
| **Version** | v7 | v2 | v2 | v4.9.0 | v3.0 |
| **Stage of development** | On the market | On the market | On the market | On the market | On the market |
| **Current market** | Used in over 40 countries in Africa, South Asia, Eastern Europe and North and South America. | Used in 9 countries worldwide in North America, Asia-Pacific and European regions. | Indonesia, Malaysia, Viet Nam, mainland China, South-East Asia | Asia-Pacific, Europe, Middle East, North Africa, the Americas. | Global, over 20 countries. |
| **Certification** | CE marked | CE marked | China NMPA-tier 2, others pending | CE marked | CE marked |
| **Online / offline product** | Online and offline | Online and offline | Online only | Online and offline | Online and offline |
| **Target Setting** | Primary health centers, public and private sector. | Primary health centers, general hospital (above primary level), teleradiology companies, public and private sector. | Primary health centers, teleradiology companies, public and private sectors. | Primary health centers, general hospital (above primary care level), teleradiology companies, public and private sector. | Primary health centers, general hospital (above primary level), teleradiology companies, public and private sector. |
| **Intended population** | Individuals 4 years and above | Approved for individuals 15 years and | Individuals 15 years and above | Individuals 14 years and above | Approved for use in individuals 15 years above. May also be |
| | | above. May also be used in individuals 12-15 years. | | | used in children and adolescents 6 years and above. |
| **Input** | DICOM from any kind of CXR machine (after validation).<br>PA CXR. | JPEG, PNG, DICOM from any kind of CXR machine.<br>PA CXR, AP CXR | DICOM from any kind of CXR machine.<br>PA CXR | DICOM from any kind of CXR machine.<br>PA CXR, AP CXR, Portable | JPEG, PNG, DICOM from any kind of CXR machine.<br>PA CXR, AP CXR, Portable |
| **Integration** | For version 7 it is possible to send the heatmap and score in a DICOM format to a picture archiving and communication system (PACS). | Possible to integrate the product with the client's legacy picture archiving and communication system (PACS) | Possible to integrate the product with client's legacy picture archiving and communication (PACS) system | Lunit INSIGHT CXR can be integrated with a legacy picture archiving and communication system (PACS) which communicates via DICOM C-Store. | Possible to integrate the product with the client's legacy Picture Archiving and Communication System (PACS) |
| **Compatibility with CXR machine models** | Any brand and model | Any brand and model | Any brand and model | Any brand and model, including portable x-ray machines | Any brand and model, including portable x-ray machines |
| **Output** | Heatmap, probability score for TB. | Heatmap, probability score for TB and other CXR findings, dichotomous output and location of | Heatmap, probability score for TB and other pulmonary findings, dichotomous outcome for TB. | Heatmap, probability score for TB and other CXR findings, dichotomous outcome for each abnormality, | Box indicating location of abnormality, probability score for TB and other CXR findings, dichotomous |
| | | each abnormality. | | location of each abnormality. | outcome for each abnormality. |
| **Training technique(s)** | Supervised deep learning (CNN, RNN) and manual feature engineering | Supervised deep learning (CNN, RNN) | Supervised deep learning (CNN, RNN) | Supervised deep learning (CNN) | Not provided |
| **Trained on** | Over 1M images from various regions and settings around the world. | Not provided | 124,394 images from China | 200,000 images | 2,500,000 images |

**Abbreviations.** CXR: Chest X-ray, PA: posterior-anterior, AP: anterior-posterior, CNN: convolutional neural network, RNN: recurrent neural network.

## Annex 4. X-ray machine specification and radiologist reading details

| **Product name: DELFT Easy DR** |
|---|
| Company: Delft Imaging, the Netherlands |
| Installation date: 03/02/2014 |
| **Functional Description** |
| Digital Radiography system with single flat panel detector, capable to take digital chest x-rays in erect position (horizontal x-ray beam) maintaining 180 cm distance from x-ray source to detector. |
| **X-ray Generator** |
| High frequency inverter technology for constant output |
| Output power 50 Kw (lower or higher power options available) |
| kV range: 40-150 kV |
| Exposure time range: 0.1-6.2 sec |
| Maximum filament current not smaller than 350mA at 100KV with AEC device |
| **X-ray Tube and Collimator** |
| High-speed rotating anode with dual focus tube, dual focus 0.6/1.2mm |
| mA range: up to 630 milliampere |
| Tube anode heating capacity: 300KHU |
| Total filtration with collimator: 1 to 3.2 mm Al |
| **X-ray Detector** |
| CANON CXDI 401-C/G series, 43 cm x 43 cm |
| Pixel Matrix: 3320x3408 |
| Pixel pitch: 125µm - 125 µm |
| A/D conversion:  16 bits minimum |
| Spatial resolution:  4 lp/mm |
| Image Preview: 3.2 sec |

Radiologist reading process: x-ray capture -> image upload to PACS system -> download as JPEG -> send to radiologists via email -> the radiologist alternated x-ray reading and send back results via email -> radiographer input results into OpenMRS and generate formal report.
Safety measures were in place for pregnant women and children. If there is no strong recommendation by a physician, we did not allow pregnant women to undergo X-ray. In case of children, we confirmed referral from a physician and permission from the parents before performing x-examination. In the cases of x-ray was required, we cautiously exposed the body parts that is precisely recommended by the physicians, along with proper counselling. A lead apron is used in case of a pregnant women to prevent any radiation scatter, the primary beam is targeted away from the pelvis. In children, the dose was reduced. For children, the default setting for chest radiographs with the Delft Easy DR X-ray System was 81 kV & 160 mA (compared to 125 kV and 200 mA for adults). Depends on body structure and weight the actual setting was slightly adjusted when needed.

## Annex 5 Python script used to anonymize images and save it with unique patient ID and associated stats

```
Tags relevant for this file:
AccessionNumber
PatientID
StudyInstanceUID
SeriesInstanceUID
SOPInstanceUID
SeriesDescription
StudyDescription
BodyPartExamined
Age
Gender
Modality
"""
import pydicom as dicom
import os
import sys
import csv
import hashlib
from tqdm import tqdm
import pandas as pd
AccessionNumber = [0x0008, 0x0050]
StudyInstanceUID = [0x0020, 0x000d]
SeriesInstanceUID = [0x0020, 0x000e]
SOPInstanceUID = [0x0008, 0x0018]
header = str.encode('/x00\x00\x00\x00\x00\x00\x00\x00\x00\x00\x00'\
'\x00\x00\x00\x00\x00\x00\x00\x00\x00\x00\x00\x00'\
'\x00\x00\x00\x00\x00\x00\x00\x00\x00\x00\x00\x00'\
'\x00\x00\x00\x00\x00\x00\x00\x00\x00\x00\x00\x00'\
'\x00\x00\x00\x00\x00\x00\x00\x00\x00\x00\x00\x00'\
'\x00\x00\x00\x00\x00\x00\x00\x00\x00\x00\x00\x00'\
'\x00\x00\x00\x00\x00\x00\x00\x00\x00\x00\x00\x00'\
'\x00\x00\x00\x00\x00\x00\x00\x00\x00\x00\x00\x00'\
'\x00\x00\x00\x00\x00\x00\x00\x00\x00\x00\x00\x00'\
'\x00\x00\x00\x00\x00\x00\x00\x00\x00\x00\x00\x00'\
'\x00\x00\x00\x00\x00\x00\x00\x00\x00DICM')
def hash_algo(value):
"""Hash the value using sha-224."""
return str(hashlib.sha224(value.encode('utf-8')).hexdigest())
# this is to create a hash, which is uniqque ranodmzely generated value based on the value

def tag_exists(dcmimg, tag):
"""Check if tag exists."""
```

```
if isinstance(tag, tuple):
if (tag[0], tag[1]) in dcmimg:
return True
elif tag in dcmimg:
return True
return False

def get_tag_value(dcmimg, tag):
"""Return tag if tag is available else returns None."""
if tag_exists(dcmimg, tag):
if isinstance(tag, tuple):
return dcmimg[tag[0], tag[1]].value
else:
return getattr(dcmimg, tag)
return None

def check_valid_dicom(dcmimg):
"""Check if the dicom is valid chest x-ray."""
def check_header(dcmimg):
data = open(dcmimg, 'rb').read()
if header in data:
return True
def check_study_description(dcmimg):
"""Check if dicom is valid or not."""
words_list = [
"mammo",
"spine",
"upper extremities",
"lower extremities"
]
study_description = get_tag_value(dcmimg, 'StudyDescription')
if study_description:
for word in words_list:
if study_description.lower().find(word) != -1:
return False
return True
if check_header(dcmimg):
return True
return False

class MedData(object):
"""Save medical data."""
def __init__(self, dicom_obj):
"""Initialize with a dicom object."""
self.PatientID = get_tag_value(dicom_obj, 'PatientID')
self.AccessionNumber = get_tag_value(dicom_obj, 'AccessionNumber')
```

```
self.StudyInstanceUID = get_tag_value(
dicom_obj, (StudyInstanceUID[0], StudyInstanceUID[1]))
self.SeriesInstanceUID = get_tag_value(
dicom_obj, (SeriesInstanceUID[0], SeriesInstanceUID[1]))
self.SOPInstanceUID = get_tag_value(
dicom_obj, (SOPInstanceUID[0], SOPInstanceUID[1]))
self.errors = []
self.hashes = {}
def add_error(self, error):
"""Add type of the error to error list."""
self.errors.append(str(type(error)))
def add_hash(self, name, val, hash_val):
"""Add hashes to the list."""
self.hashes[name]= val
self.hashes[name+' hash'] = hash_val

def get_tags(dicom_tags_path):
"""Get tags from csv file in path and anonymize or hash as mentioned."""
dicom_list = []
with open(dicom_tags_path, 'r') as file_obj:
reader = csv.DictReader(file_obj, delimiter=',')
for line in reader:
dicom_tag_tuple = line['DicomTags'].strip().split(' ')[0]
dicom_tag_tuple = dicom_tag_tuple[1:-1]
dicom_tag_ids = dicom_tag_tuple.split(",")
dicom_tag_ids = [hex(int(id, 16)) for id in dicom_tag_ids]
dicom_tag_ids.append(0)
dicom_tag_ids.append(0)
if line['Hashing'].strip() == 'Yes':
dicom_tag_ids[3] = 1
elif line['Anonymisation Required'].strip() == 'Yes':
dicom_tag_ids[2] = 1
dicom_list.append(dicom_tag_ids)
return dicom_list

def anonymize(dicom_tags_list):
"""Return a function to anonymize as per dicom_tags_list."""
def run(fname):
try:
dcmimg = dicom.read_file(fname,force=True)
"""Take images which are valid chest x-rays."""
med_data_obj = MedData(dcmimg)
for tag in dicom_tags_list:
try:
if not tag_exists(dcmimg, (tag[0], tag[1])):
# print(tags)
```

```
continue
if tag[3] == 1:
value = dcmimg[tag[0], tag[1]].value
hashed_val = hash_algo(value)
dcmimg[tag[0], tag[1]].value = hashed_val
med_data_obj.add_hash(dcmimg[tag[0], tag[1]].name, value, hashed_val)
elif tag[2] == 1:
dcmimg[tag[0], tag[1]].value = ''
except Exception as e:
dcmimg[tag[0], tag[1]].value = ''
med_data_obj.add_error(e)
return dcmimg, med_data_obj
except Exception as e:
# raise e
return None, str(type(e))
return run

def create_anonymized_xray_data(input_path, output_path, csv_path, anonymizer):
"""Create anonymized xray data, selecting images from input_path and writing in output_path."""
data_details = []
accession_number_not_available = []
error_data = []
hashes_data = {}
num_files = 0
for root, dirs, files in os.walk(input_path):
if len(files) == 0:
continue
filelist = []
for filename in files:
filepath = os.path.join(root, filename)
filelist.append(filepath)
for fname in tqdm(filelist):
dcmimg, med_data_obj = anonymizer(fname)
if dcmimg is None:
if med_data_obj is not None:
error_data.append([fname, '', med_data_obj])
continue
num_files += 1
filename = None
filename = os.path.basename(fname)
outname = os.path.join(output_path, filename)
data_row = [med_data_obj.PatientID, med_data_obj.AccessionNumber,
med_data_obj.StudyInstanceUID]
dicom.write_file(outname, dcmimg)
# for hashes in med_data_obj.hashes:
# hashes_data.append(hashes)
```

```
hashes_data[filename]= med_data_obj.hashes
if (get_tag_value(dcmimg, 'AccessionNumber') is None or
len(str(get_tag_value(dcmimg, 'AccessionNumber'))) == 0):
accession_number_not_available.append([fname])
if data_row not in data_details:
data_details.append(data_row)
"""Create a csv."""
# csvpath = os.path.join(csv_path, 'PatientData.csv')
# create_csv(csvpath, data_details)
"""CSV for accession number not available."""
acc_not_avail_path = os.path.join(csv_path, 'AccessionNotAvailableData.csv')
create_csv(acc_not_avail_path, accession_number_not_available)
"""CSV for error."""
error_path = os.path.join(csv_path, 'Errors.csv')
create_csv(error_path, error_data)
"""CSV for hashes."""
hashes_path = os.path.join(csv_path, 'Hashes.csv')
# create_csv(hashes_path, hashes_data)
hashes_df = pd.DataFrame.from_dict(hashes_data,orient='index')
hashes_df.index.name = 'Filename'
hashes_df.to_csv(hashes_path)

print("Total number of chest x-rays found are {}".format(num_files))

def create_csv(csvpath, data):
"""Create a csv from the data."""
with open(csvpath, "w+") as csvfile:
csvwriter = csv.writer(csvfile)
for datum in data:
csvwriter.writerow(datum)

def calling_main_function():
# dicom_tags_path = sys.argv[1]
dicom_tags_path = "xxxxxxxxxx"
dicom_tags_list = get_tags(dicom_tags_path)
# input_path = sys.argv[2]
input_path = "xxxxxxxxxx"
# output_path = sys.argv[3]
output_path = "xxxxxxxxxx"
# csv_path = sys.argv[4]
csv_path = "xxxxxxxxxx"
create_anonymized_xray_data(input_path, output_path,csv_path, anonymize(dicom_tags_list))

if __name__ == '__main__':
calling_main_function()
```

**For the source code used in this analysis please refer to our GitHub repository at**
https://github.com/ZZQin/BGD_AI_for_TB_Detection

## Annex 6 Binary classifications derived from radiologist's categories

| | | | | |
|---|---|---|---|---|
| Radiologist's Category | *Highly Suggestive of TB* | *Possibly TB* | *Abnormal, but not TB* | *Normal* |
| Binary classification A | *Radiologically Positive* | *Radiologically Negative* | | |
| Binary classification B | *Radiologically Positive* | | *Radiologically Negative* | |
| Binary classification C | *Radiologically Positive* | | | *Radiologically Negative* |

## Annex 7 Comparing ROC curve and Precision Recall curve

The ROC curve is often used to illustrate the performance of binary classifiers (like a diagnostic test); alternative measurements like positive predictive value (PPV) and the PRC are less commonly used. Binary classifiers can be applied to data sets that are strongly imbalanced-where the number of true negatives is much larger than the number of true positives. Saito (2015) plot a ROC and PRC for a binary classifier when applied to both a balanced (10 positives/10 negatives) and imbalanced (5 positives/15 negatives) dataset. Calculating the AUC for both the balanced and imbalanced ROC and PRC curves, they found the AUC (ROC) did not reflect the poor performance of the classifier on the imbalanced dataset, but the PRAUC did. The PRC's baseline changes based on the number of positive and negatives in a dataset, so the PRAUC changes accordingly, the ROC's baseline does not consider change in this way. Saito (2015)'s findings are reinforced by their literature review and re-analysis on an example dataset.

## Annex 8 p values when comparing AUC between AI products

| **AI Product** | **AUC (95% CI)** | **p value** | | | | |
|---|---|---|---|---|---|---|
| | | **CAD4TB** | **InferRead DR** | **JF-CXR-1** | **Lunit INSIGHT CXR** | **qXR** |
| **CAD4TB** | 90.34% (89.81-90.87%) | | $< 2.2e^{-16}$ | $< 2.2e^{-16}$ | $< 2.2\ e^{-16}$ | 0.00053 |
| **InferRead DR** | 84.90% (95% CI: 84.27-85.54%) | | | **0.97** | $< 2.2\ e^{-16}$ | $< 2.2e^{-16}$ |
| **JF-CXR-1** | 84.89% (95% CI: 84.26-85.53%) | | | | $< 2.2e^{-16}$ | $< 2.2e^{-16}$ |
| **Lunit INSIGHT CXR** | 88.61% (95% CI: 88.03-89.20%) | | | | | $< 2.2e^{-16}$ |
| **qXR** | 90.81% (95% CI: 90.33-91.29%) | | | | | |

# Annex 9 AI product outputs when reading the same CXR image.

Heatmap from CAD4TB

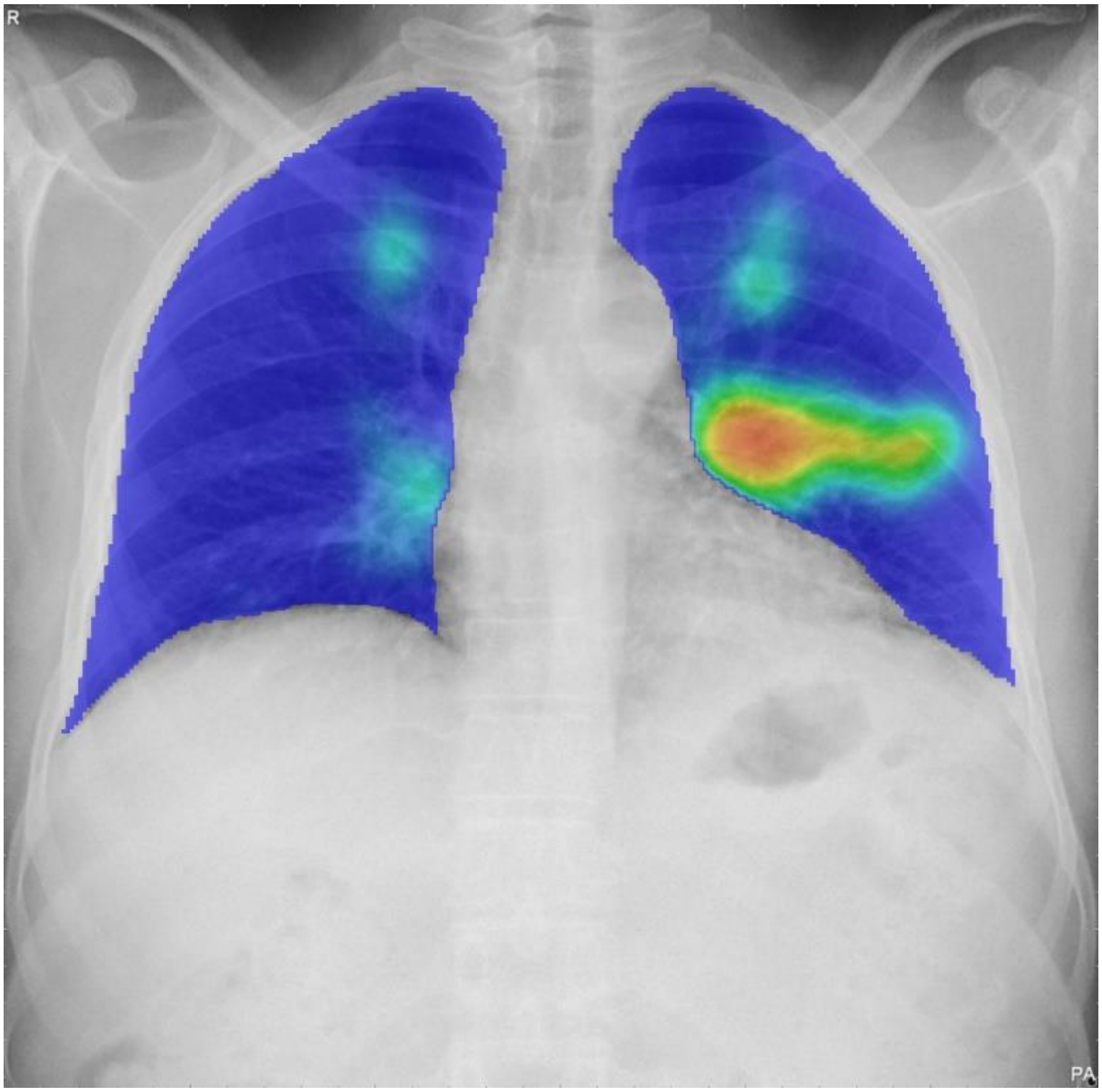

Heatmap from InferRead

InferVision

Image exam report:

| | | |
|---|---|---|
| Patient ID number: | Examination number: | Examination date: |
| Name: | Gender: | Age: |

Modality: DX
Examination type: Chest X-ray Examination

Medical image:

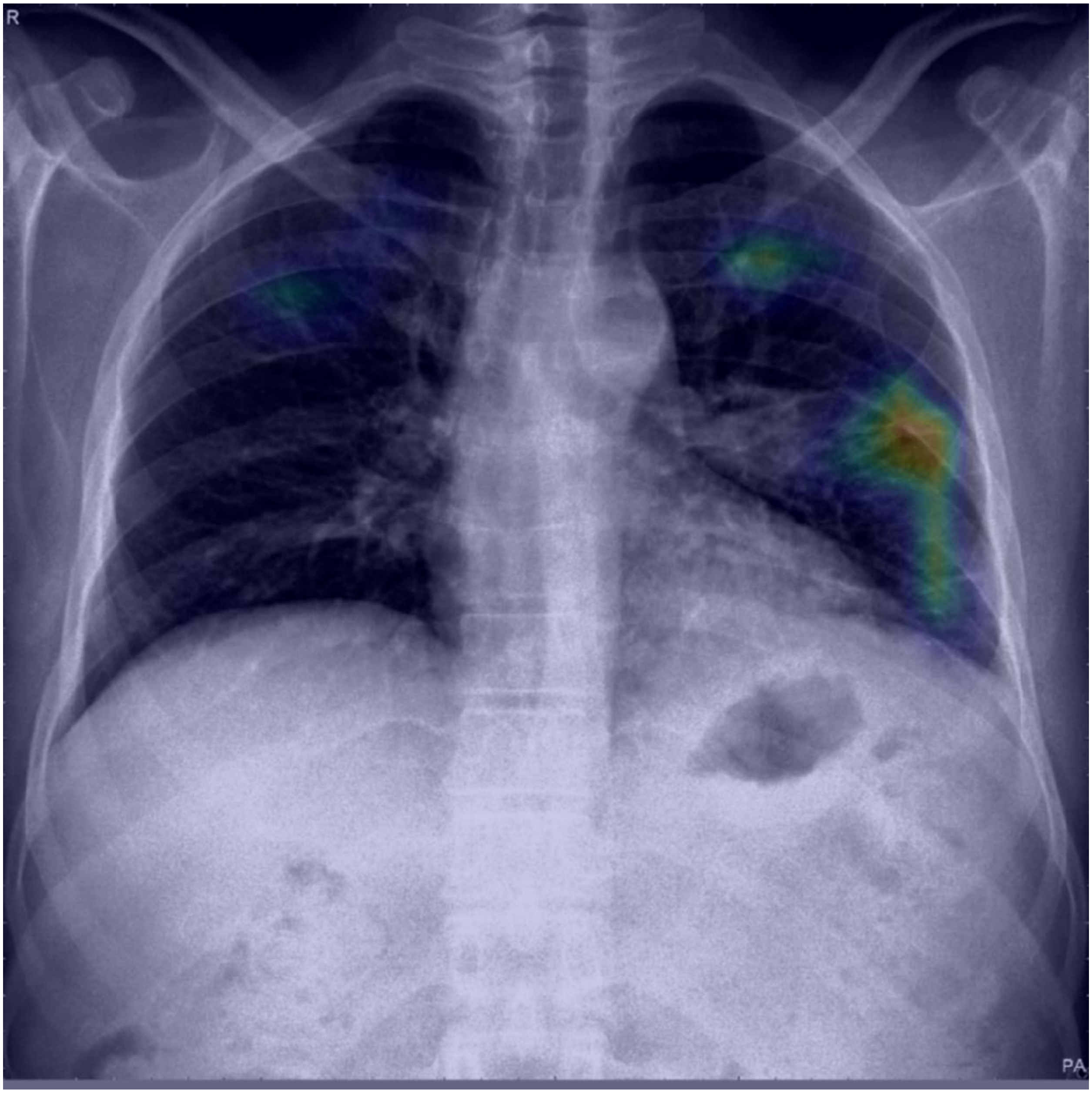
R
PA

Heatmap from Lunit INSIGHT CXR

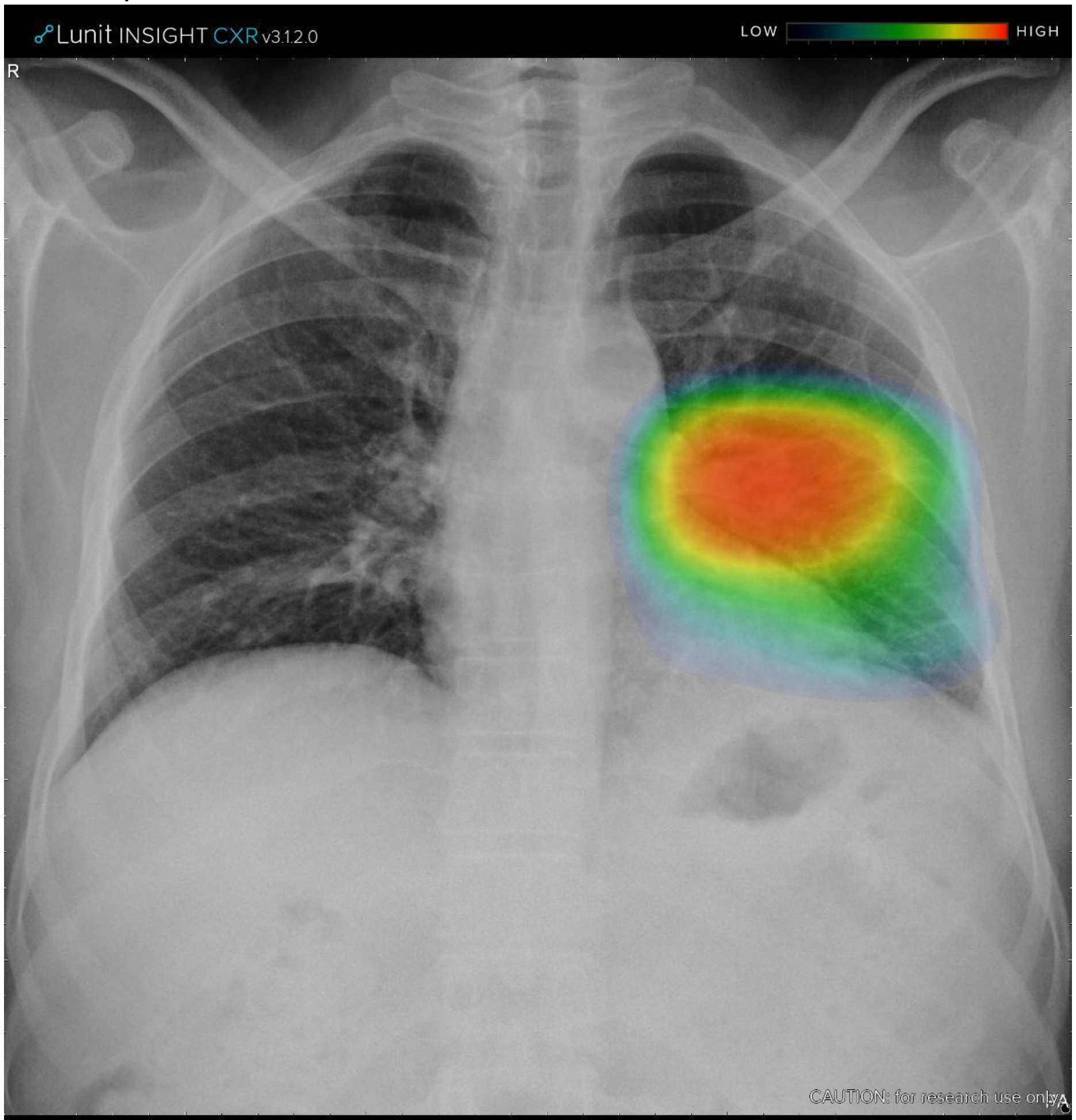

CASE REPORT

| | | | |
|---|---|---|---|
| Abnormality Score 94% | | TB Screening Score 94% | |
| Atl | Atelectasis | 38% | Multiple Lesions |
| Calc | Calcification | Low | - |
| Cm | Cardiomegaly | Low | - |
| Csn | Consolidation | 94% | Left Upper zone |
| Fib | Fibrosis | Low | - |
| MW | Mediastinal Widening | Low | - |
| Ndl | Nodule | 41% | Left Upper zone |
| PEf | Pleural effusion | Low | - |
| Ppm | Pneumoperitoneum | Low | - |
| Ptx | Pneumothorax | Low | - |

Heatmap from qXR

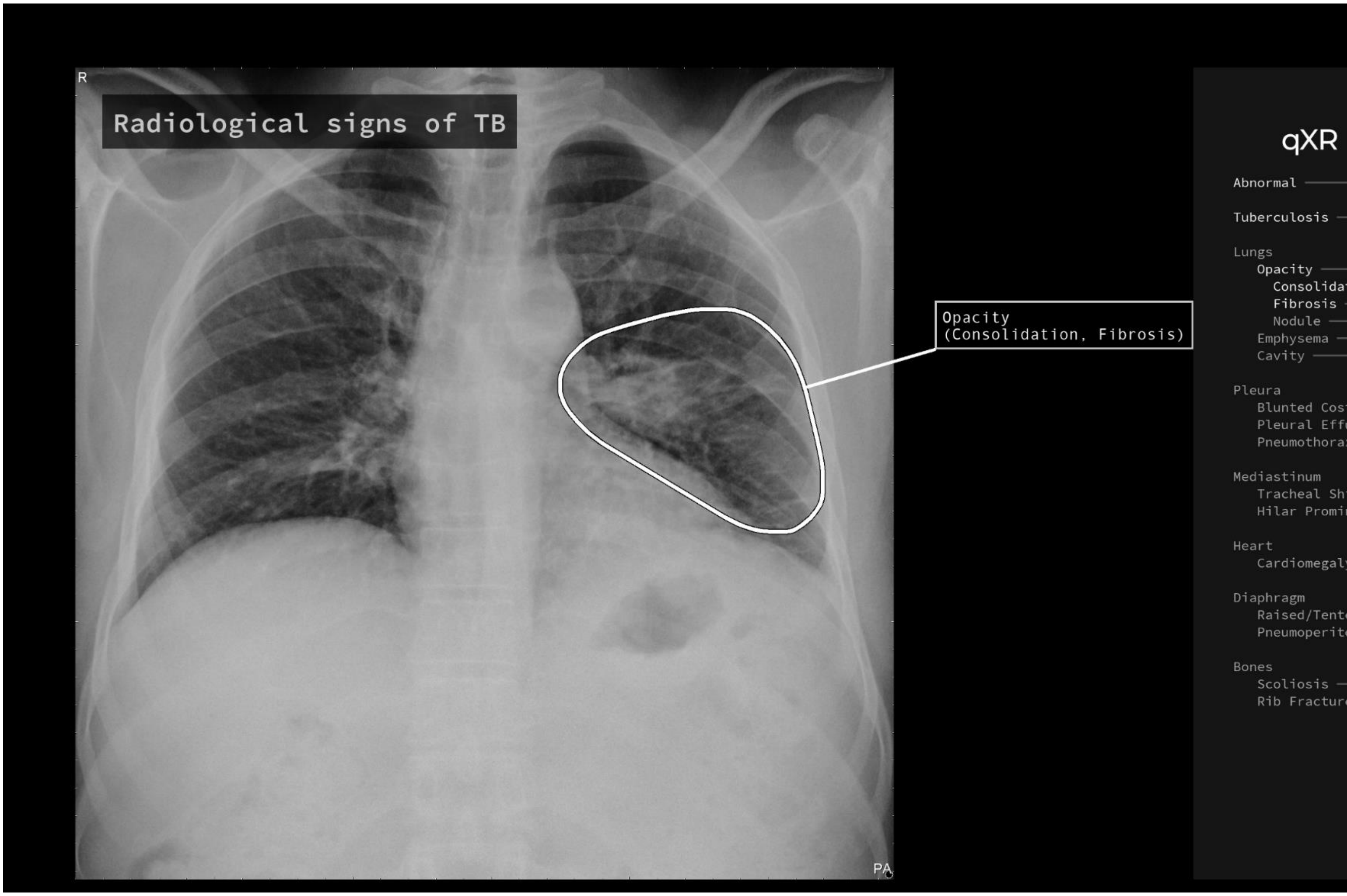

Annex 10 Stacked density plot of the distributions of the abnormality scores of five AI algorithms disaggregated by Xpert outcomes and by prior TB history. The dark and light red bars were Bac negative and the dark and light green bars were Bac positive.

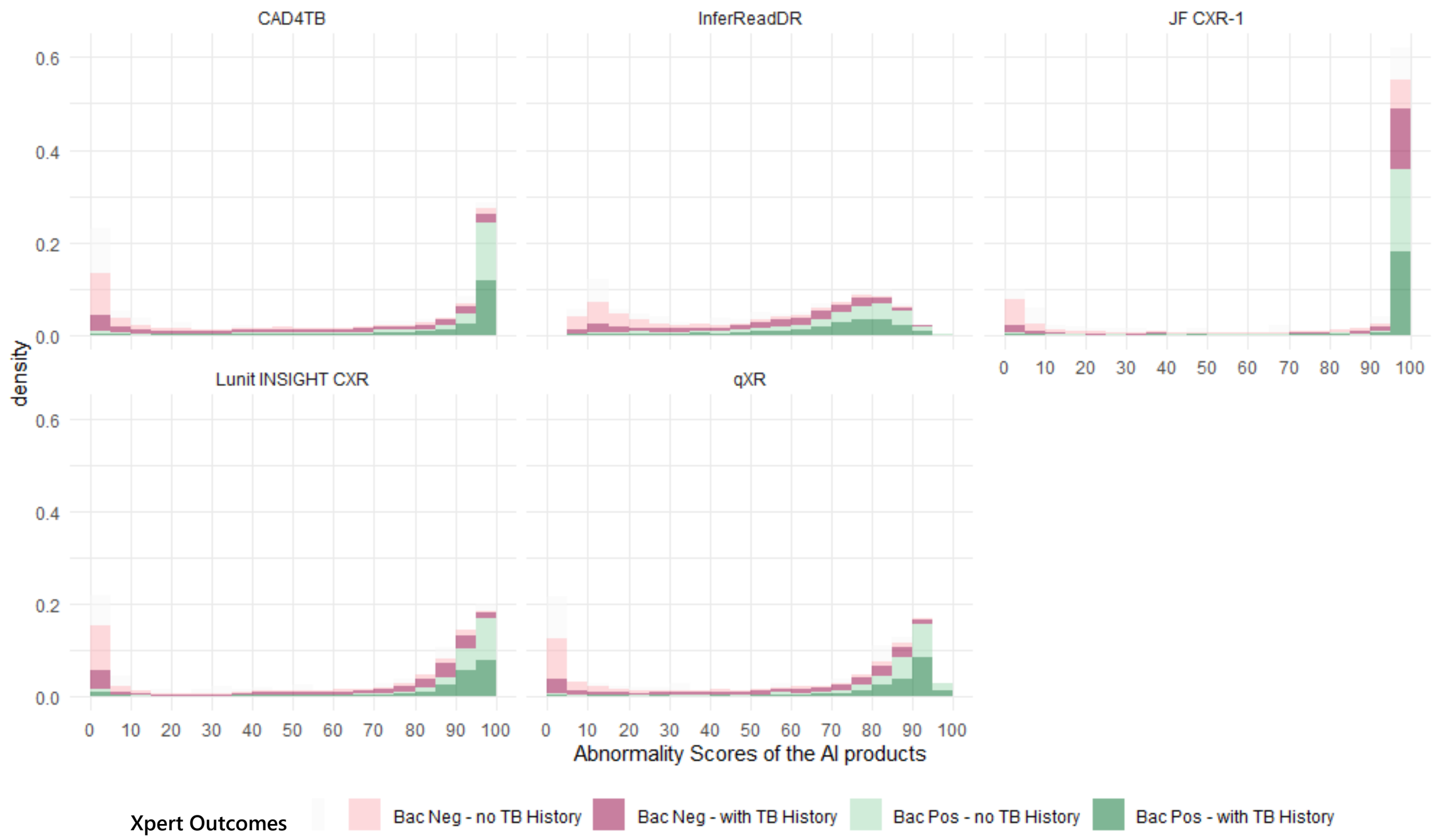

## Annex 11 The p-value of the AUC of the five CAD products in different subgroups stratified by age, patient source, prior TB history and gender

| **CAD4TB** | Middle age | Old age |
|---|---|---|
| Young age | 0.0415434 | 4.326E-11 |
| Middle age | | 6.75E-10 |
| Old age | | |

| **qXR** | Middle age | Old age |
|---|---|---|
| Young age | 0.0032515 | 1.835E-16 |
| Middle age | | 6.732E-13 |
| Old age | | |

| **Lunit INSIGHT CXR** | Middle age | Old age |
|---|---|---|

| **CAD4TB** | Previously treated cases |
|---|---|
| New cases | 1.1305E-12 |
| Previously treated cases | |

| **qXR** | Previously treated cases |
|---|---|
| New cases | 6.7377E-15 |
| Previously treated cases | |

| **Lunit INSIGHT CXR** | Previously treated cases |
|---|---|
| New cases | 1.6411E-18 |
| Previously treated cases | |

| **CAD4TB** | Male |
|---|---|
| Female | 0.799505 |
| Male | |

| **qXR** | Male |
|---|---|
| Female | 0.382043 |
| Male | |

| **Lunit INSIGHT CXR** | Male |
|---|---|
| Female | 0.019814 |
| Male | |

| | | |
|---|---|---|
| Young age | 0.2367085 | 9.577E-08 |
| Middle age | | 6.662E-08 |
| Old age | | |

| **JF CXR-1** | Middle age | Old age |
|---|---|---|
| Young age | 0.005172 | 6.718E-14 |
| Middle age | | 3.425E-11 |
| Old age | | |

| **InferReadDR** | Middle age | Old age |
|---|---|---|
| Young age | 0.0032294 | 7.16E-14 |
| Middle age | | 2.038E-10 |
| Old age | | |

| **JF CXR-1** | Previously treated cases |
|---|---|
| New cases | 1.3367E-23 |
| Previously treated cases | |

| **InferReadDR** | Previously treated cases |
|---|---|
| New cases | 1.5936E-30 |
| Previously treated cases | |

| **JF CXR-1** | Male |
|---|---|
| Female | 0.004484 |
| Male | |

| **InferReadDR** | Male |
|---|---|
| Female | 0.632632 |
| Male | |

| **CAD4TB** | Public referral | DOTS retested | Walk-In | Community | Contacts |
|---|---|---|---|---|---|
| Private referral | 0.2209 | 0.057521 | 3.79E-18 | 0.922857 | 0.885101 |
| Public referral | | 0.793989 | 0.017135 | 0.694277 | 0.851686 |
| DOTS retested | | | 2.63E-06 | 0.764312 | 0.908629 |
| Walk-In | | | | 0.139937 | 0.377596 |
| Community | | | | | 0.945152 |
| Contacts | | | | | |

| **qXR** | Public referral | DOTS retested | Walk-In | Community | Contacts |
|---|---|---|---|---|---|
| Private referral | 0.87363 | 0.000208 | 1.84E-14 | 0.819641 | 0.858888 |
| Public referral | | 0.186874 | 0.002126 | 0.886959 | 0.83433 |
| DOTS retested | | | 6.87E-04 | 0.634959 | 0.595326 |
| Walk-In | | | | 0.185485 | 0.335991 |
| Community | | | | | 0.793642 |
| Contacts | | | | | |

| **Lunit INSIGHT CXR** | Public referral | DOTS retested | Walk-In | Community | Contacts |
|---|---|---|---|---|---|
| Private referral | 0.31716 | 0.081584 | 1.33E-08 | 0.736796 | 0.707993 |
| Public referral | | 0.885464 | 0.065868 | 0.489871 | 0.500426 |
| DOTS retested | | | 1.08E-03 | 0.503349 | 0.515715 |
| Walk-In | | | | 0.126034 | 0.172999 |
| Community | | | | | 0.939375 |
| Contacts | | | | | |

| **JF CXR-1** | Public referral | DOTS retested | Walk-In | Community | Contacts |
|---|---|---|---|---|---|
| Private referral | 0.90977 | 2.96E-06 | 7.48E-25 | 0.538046 | 0.35168 |
| Public referral | | 0.051939 | 3.3E-05 | 0.556622 | 0.367768 |
| DOTS retested | | | 1.07E-05 | 0.563155 | 0.954214 |
| Walk-In | | | | 0.045237 | 0.321601 |
| Community | | | | | 0.692491 |
| Contacts | | | | | |

| **InferReadDR** | Public referral | DOTS retested | Walk-In | Community | Contacts |
|---|---|---|---|---|---|
| Private referral | 0.70482 | 0.003914 | 2.09E-19 | 0.548551 | 0.871706 |
| Public referral | | 0.13896 | 5.87E-05 | 0.733783 | 0.942238 |
| DOTS retested | | | 6.56E-07 | 0.21465 | 0.665029 |
| Walk-In | | | | 0.006501 | 0.288864 |
| Community | | | | | 0.937484 |
| Contacts | | | | | |

p-values less than 0.05 are highlighted in red, showing statistical significance.